\documentclass[12pt]{article}
\usepackage{physics}
\usepackage{epsf}
\usepackage{aas_macros}
\usepackage{amssymb}
\usepackage{comment}
\usepackage{amsmath}
\usepackage{braket}
\usepackage{mathrsfs}
\usepackage{mathtools}
\usepackage{array}
\usepackage{fancyhdr}
\usepackage[dvipdfmx]{graphicx}
\usepackage{color}
\usepackage{cite}
\usepackage{bm}
\usepackage{url}
\usepackage{fancyhdr}
\usepackage[colorlinks,citecolor=blue]{hyperref}
\usepackage[hang,small,bf]{caption}
\usepackage[subrefformat=parens]{subcaption}
\renewcommand{\thefootnote}{\#\arabic{footnote}}

\renewcommand{\thefootnote}{\fnsymbol{footnote}}
\def\thefootnote{\fnsymbol{footnote}}

\makeatletter

\@addtoreset{equation}{section}
\makeatother

\def\be{\begin{equation}}
\def\ee{\end{equation}}
\def\ben{\begin{eqnarray}}
\def\een{\end{eqnarray}}

\usepackage{scalerel}
\usepackage[usestackEOL]{stackengine}

\def\dashint{\,\ThisStyle{\ensurestackMath{%
            \stackinset{c}{.2\LMpt}{c}{.5\LMpt}{\SavedStyle-}{\SavedStyle\phantom{\int}}}%
        \setbox0=\hbox{$\SavedStyle\int\,$}\kern-\wd0}\int}

\begin{document}

%\begin{titlepage}

\begin{center}

\vskip .75in

{\Large \bf On the Phase-Magnitude Relation in Gravitational Lensing: Reformulation and Applications of the Kramers-Kronig relation}

\vskip .75in

{\large
Teruaki Suyama$\,^1$, Shasvath J. Kapadia$\,^2$
}

\vskip 0.25in

{\em
$^{1}$Department of Physics, Institute of Science Tokyo, 2-12-1 Ookayama, Meguro-ku,
Tokyo 152-8551, Japan \\
$^{2}$Inter-University Centre for Astronomy and Astrophysics, Post Bag 4, Ganeshkhind, Pune 411007, India
}

\end{center}
\vskip .5in

\begin{abstract}
It is known that the amplification factor, defined as the ratio of the 
lensed to the unlensed waveform in the frequency domain,
satisfies the Kramers-Kronig (KK) relation, 
which connects the real and imaginary parts of the amplification factor for any lensing signal. 
In this work, we reformulate the KK relation in terms of the magnitude 
and phase of the amplification factor. 
Unlike the original formulation, the phase cannot be uniquely determined 
from the magnitude alone due to the possible presence of a Blaschke product. 
While this ambiguity does not arise in the case of a point-mass lens, 
it can appear in more complex lens models, 
such as those with an NFW lens profile.
As an application of our formulation, we demonstrate that the leading-order 
behavior of the phase in the low-frequency regime is completely determined 
by the leading-order behavior of the magnitude in the same regime.
This reproduces known results from the literature, derived via 
low-frequency expansions for specific lens models. 
Importantly, our result does not rely on any particular lens model, 
highlighting a universal feature that the low-frequency behavior of the 
amplification factor is tightly constrained by the KK relation.
As a further application, we present two examples in which the phase 
is constructed from a given analytic form of the magnitude 
using the newly derived KK relation. 
In particular, the second example allows for an analytic evaluation 
of the KK integral, yielding an explicit expression for the phase. 
This study offers a potentially powerful method for applying the KK relation 
in model-agnostic searches for lensing signals.
\end{abstract}

\renewcommand{\thepage}{\arabic{page}}
\setcounter{page}{1}
\renewcommand{\thefootnote}{\#\arabic{footnote}}
\setcounter{footnote}{0}

\section{Introduction}
Gravitational lensing of gravitational waves (GWs) promises to serve as an important tool
for probing the dark sector of the universe. 
A central quantity in gravitational lensing is the amplification factor $F(\omega)$, 
which encodes features of the lens on the waveform in the frequency domain. 
In the case of GWs, there are two natural regimes within which GW lensing may be studied. In the geometric optics regime, wherethe  Schwarzschild radius of the lens is much larger than the wavelength of the GWs, multiple temporally resolved images will be produced. Each image differs from the source GWs by a constant magnification, as well as a constant additive phase factor called the Morse phase. On the other hand, if the Schwarzschild radius of the lens is comparable to the wavelength of the GWs, then wave-optics effects will be present, and a single image modulated with respect to the unlensed GWs will be produced. Searches for lensing in the geometric and wave-optics regimes, in LIGO-Virgo-Kagra data \cite{LIGOScientific:2014pky, TheVirgo:2014hva, KAGRA:2020tym}, have been conducted. To date, no confirmed detection has been reported \cite{LIGOScientific:2018mvr,LIGOScientific:2020ibl,LIGOScientific:2021usb,KAGRA:2021vkt}. 

In a previous study~\cite{Tanaka:2023mvy}, it was shown that $F(\omega)$ satisfies a 
Kramers-Kronig (KK) relation which is a relation between the real part and the imaginary part of $F(\omega)$. 
This relation holds solely by virtue of analyticity and boundedness of $F(\omega)$
in the upper-half complex frequency plane, 
which follows from the causal nature of gravitational lensing. 
Since the KK relation is independent of specific lens models or properties of the GW source, 
it may serve as a powerful, model-independent diagnostic to probe for lensing signatures in 
observed GWs. 
Indeed, under the ideal situation where detector noise is negligible, 
it has been demonstrated that the KK relation can be used to rule out spurious lensing features 
that are inconsistent with causality \cite{Tanaka:2025ntr}.

In this paper, we reformulate the original KK relation, 
which relates the real and imaginary parts of $F(\omega)$,
into a relation between the magnitude $|F(\omega)|$ and the phase $\theta(\omega)$. 
Unlike the real and imaginary parts, which are rather mathematical quantities without immediate physical interpretation, 
the magnitude and phase have clear physical meaning. This motivates us to formulate a new KK relation expressed in terms of $|F(\omega)|$ and $\theta(\omega)$. 
In fact, a similar transformation has proven extremely useful in condensed matter physics. 
A KK relation between the magnitude and phase refers to the optical reflectivity and
measurements of reflectivity by experiments, and combining it with the KK relation enables 
extraction of optical constants of the material under consideration ~\cite{STERN1963299}. 
However, as we will show, due to the differences in physical context, the gravitational lensing case may include additional terms that are absent in condensed matter systems. 
This leads to an interesting possibility that two distinct lensing systems may exhibit the same $|F(\omega)|$ 
but differ in the phase $\theta(\omega)$. 
We do not further explore such degeneracies in this paper, and
it remains unclear whether astrophysical lenses exhibit such degeneracies. 

As an interesting application of the newly derived KK relation, 
we investigate the low-frequency behavior of $F(\omega)$. 
Previous studies have shown that, for some specific lens models,
the leading-order frequency dependence of $F(\omega)$ at low frequencies is dictated 
by the slope of the lens's density profile~\cite{Choi:2021bkx, Tambalo:2022plm}. 
We show that even without assuming any particular lens model,
the frequency dependence of the phase $\theta(\omega)$ at low frequencies 
is completely determined once $|F(\omega)|$ is known in that regime, as required by the KK relation.
This indicates that the fundamental principle of causality imposes strong constraints 
on the behavior of $F(\omega)$ in the low-frequency limit. 
%\sk{I did not fully understand this paragraph. If $F(\omega)$ is already determined by the slope of the density profile, then so is $\theta(\omega)$. My understanding was that $\theta(\omega)$ is fixed once $|F(\omega)|$ is known.}

Another application is the construction of phenomenological models of the amplification factor. 
A recent work~\cite{Chakraborty:2024mbr}, inspired by the shape of the amplification factor
for some representative lens models, 
proposed a phenomenological functional form for $|F(\omega)|$ which contains several fitting parameters, 
without assuming a specific lens configuration. 
Although the phase was not specified in that construction, 
our results imply that the phase cannot be assigned independently. 
Instead, it is constrained by the KK relation once the magnitude is given. 
In principle, this result should be accounted for in searches for 
gravitational lensing signals in GW data based on phenomenologically assumed lens signal.
We also provide an example of simple analytic form of the magnitude of the
amplification factor for which the corresponding phase obtained by the KK relation
is also written in terms of known simple functions.

\section{KK Relations between magnitude and phase}
\label{Sec:Derivation-KK}
\subsection{Mathematical preliminaries}
This subsection provides an overview of the mathematical derivation of KK relations between magnitude and phase of an analytical function based on \cite{STERN1963299}.

In general, a complex function $A(\omega)$ satisfies the KK relation,
which provides a constraint between the real and imaginary parts,
if it is analytic in the upper-half complex $\omega$-plane (we denote it by $I_+$) and bounded as $|\omega| \to \infty$. 
On the other hand, analyticity and boundedness do not uniquely determine the phase 
of $A(\omega)$ from the magnitude.
This can be seen by defining a new function $A_{\rm new}(\omega)$ as \cite{STERN1963299}:
\be
\label{def-A-new}
A_{\rm new}(\omega)=B(\omega) A_(\omega),
\ee
where $B(\omega)$ is the Blaschke product defined by:
\be
\label{def:Blaschke}
B(\omega)=\prod_n \left( \frac{\omega-\mu_n}{\mu_n^*-\omega} \right),
\ee
where $\mu_n$ are (arbitrary) complex constants with non-negative imaginary parts.
To satisfy the reality condition $A^*(\omega)=A(-\omega)$ for $\omega \in \mathbb{R}$,
each $\mu_n$ must be paired with a corresponding $\mu_m$ such that their real parts have opposite signs while their imaginary parts have the same sign.
Then, it is immediate to show that $|B(\omega)|=1$ for $\omega \in \mathbb{R}$, from which
it follows $|A_{\rm new}(\omega)|=|A(\omega)|$.
Since the Blaschke product is analytic and bounded in $I_+$,
$A_{\rm new}(\omega)$ is also analytic and bounded in the same region.
Thus, two functions, which are analytic and bounded in $I_+$
and have the same magnitude on $\mathbb{R}$, have different phase as determined by the
Blaschke product.

\begin{figure}[t]
  \begin{center}
    \includegraphics[clip,width=11.0cm]{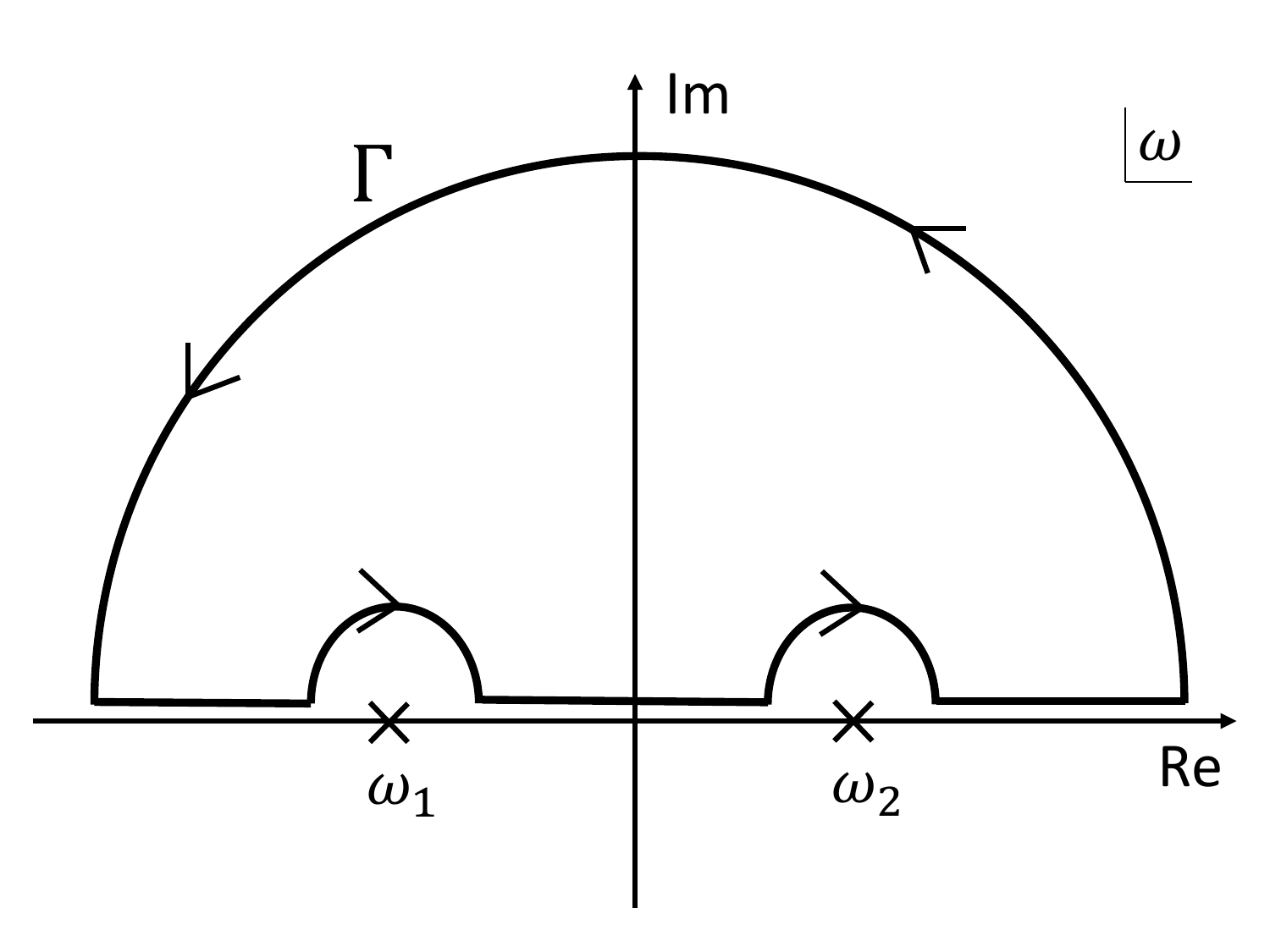}
    \caption{Contour $\Gamma$ of integral used to derive the KK relation
    for magnitude and phase.}
    \label{fig:kramers-kronig}
  \end{center}
\end{figure}

Having this in mind, let us suppose that a comlex function $A(\omega)$, which is
analytic and bounded, has no zero-points (points where $A(\omega)=0$) in $I_+$.
Then $\ln A(\omega)$ has no branch points and is analytic and bounded,
ensuring that $\ln A(\omega)$ also satisfies the KK relation as $A(\omega)$ does.
To write down its explicit form, let us define a new function $G(\omega)$ by
\be
G(\omega)=\left( \frac{1}{\omega-\omega_1}-\frac{1}{\omega-\omega_2} \right) \ln A(\omega),
\ee
where $\omega_1, \omega_2 \in \mathbb{R}$.
This function is analytic and decays as ${\cal O}(\omega^{-2})$ for $|\omega| \to \infty$ in $I_+$.
Then, from Cauchy's theorem, the integral of $G(\omega)$ along the
contour $\Gamma$ shown in Fig.~\ref{fig:kramers-kronig} becomes zero:
\be
\int_\Gamma G(\omega') d\omega' =0.
\ee

This integral consists of three components: the one along the real axis,
the one along the small semi-circles around $\omega_1, \omega_2$, and
the one along the upper semi-circle.
The last one vanishes when the radius of the semi-circle is taken to be infinite.
Then, taking the limit where the radius of the small semi-circles is zero, the
above equation yields:
\be
\dashint_{-\infty}^\infty \left( \frac{1}{\omega'-\omega_1}-\frac{1}{\omega'-\omega_2} \right) \ln A(\omega') d\omega'+\pi i (\ln A(\omega_2)-\ln A(\omega_1) )=0,
\ee
where $\dashint$ stands for the Cauchy principal value.
Substituting the decomposition $A(\omega)=|A(\omega)|e^{i\theta (\omega)}$ into
the above relation, the real and the imaginary part of the relation become:
\begin{align}
&\dashint_{-\infty}^\infty \left( \frac{1}{\omega'-\omega_1}-\frac{1}{\omega'-\omega_2} \right) \ln |A(\omega')| d\omega'
-\pi (\theta (\omega_2)-\theta (\omega_1) )=0, \\
&\dashint_{-\infty}^\infty \left( \frac{1}{\omega'-\omega_1}-\frac{1}{\omega'-\omega_2} \right) \theta (\omega') d\omega'
+\pi (\ln |A(\omega_2)|-\ln |A(\omega_1)| )=0. \label{magnitude-from-phase}
\end{align}
In particular, choosing $\omega_1=-\omega_2=\omega$ and using the identities 
$|A(\omega)|=|A(-\omega)|, ~\theta (\omega)=-\theta (-\omega)$, the first relation becomes:
\be
\label{phase-from-magnitude}
\theta (\omega) =-\frac{2\omega}{\pi} \dashint_0^\infty
\frac{\ln |A(\omega')|}{\omega'^2-\omega^2}d\omega',
\ee
which determines the phase $\theta (\omega)$ in terms of the magnitude $|A(\omega)|$.

Next, consider the case where $A(\omega)$ has zero points in $I_+$ \cite{STERN1963299}.
Then, based on the discussion on the Blaschke product after Eqs.~(\ref{def-A-new}),
the phase receives a contribution from the Blaschke product and the
relation between the magnitude and the phase is modified as:
\be
\label{mag-phase:relation-with-B}
\theta (\omega) =-\frac{2\omega}{\pi} \dashint_0^\infty
\frac{\ln |A(\omega')|}{\omega'^2-\omega^2}d\omega'-i \ln B(\omega).
\ee
Using Eq.~(\ref{def:Blaschke}), the phase shift due to the Blaschke product is written as:
\be
-i \ln B(\omega)=2 \sum_n \tan^{-1} \left( \frac{\omega-{\Re \mu_n}}{\Im \mu_n} \right).
\ee
In particular, the total change of the phase from $\omega=0$ to $\omega=\infty$ becomes:
\be
-i \Delta \ln B \equiv -i \ln B(\infty)+i\ln B(0)= (2p+q)\pi,
\ee
where $p$ and $q$ represent the number of zero points 
with $\Re \mu_n \neq 0$ and $\Re \mu_n = 0$, respectively. 
While Eq.~(\ref{phase-from-magnitude}) is modified as Eq.~(\ref{mag-phase:relation-with-B}) in the
presense of the Blaschke product, such modification does not arise for the relation 
(\ref{magnitude-from-phase}) as the following identity:
\be
\dashint_{-\infty}^\infty \left( \frac{1}{\omega'-\omega_1}-\frac{1}{\omega'-\omega_2} \right) 
\ln \left( \frac{\omega'-\mu_n}{\mu_n^*-\omega'} \right) d\omega'=0
\ee
holds.

\subsection{KK relations for the amplification factor}
The discussion in the previous subsection is purely mathematical and applies to any physical system, provided the relevant function satisfies analyticity and boundedness conditions.
In condensed matter physics, for instance, the normal optical reflectivity $r(\omega)$, 
derived from the complex dielectric constant, asymptotically behaves as
$r(\omega) \simeq \frac{1}{4}\omega_p^2/\omega^2$ ($\omega_p$ is the plasma frequency) at high frequency.
Substituting this asymptotic form into Eq.~(\ref{mag-phase:relation-with-B}) and 
noting that the phase approaches $\pi$ in the high-frequency limit, 
one finds that the contribution from the Blaschke product vanishes for the normal optical reflectivity.

We now turn to gravitational lensing and examine the amplification factor $F(\omega)$.
Owing to the causal nature of the lensing process \cite{Suyama:2020lbf}, 
$F(\omega)$ is known to be analytic and bounded in the upper-half complex plane \cite{Tanaka:2023mvy}.
However, in contrast to the condensed matter case, 
we show that the Blaschke product can contribute nontrivially to $F(\omega)$
depending on the detailed structure of the lens profile.
To see this, we consider the high-frequency limit in which the geometrical optics formulation gives 
the correct form of the amplification factor:
\be
F_{\rm geo}(\omega)=\sqrt{\mu_1}+\sum_{j=2}^N \sqrt{\mu_j}
e^{i\omega \Delta t_j-\pi i n_j {\rm sgn} (\omega)},
\ee
where $\mu_j$ is the magnification of the $j$-th image, $\Delta t_j$ is the time-delay of the $j$-th image compared to the first image, $e^{-i\pi n_j {\rm sgn}(\omega)}$ 
is the Morse phase,
and $N$ is the number of images.
Then, the high-frequency limit of the relation (\ref{mag-phase:relation-with-B}) becomes:
\begin{align}
\label{relation-high-f}
&\arg (\sqrt{\mu_1}+\sum_{j=2}^N \sqrt{\mu_j}
e^{i\omega \Delta t_j-\pi i n_j {\rm sgn} (\omega)}) \nonumber \\
&=-\frac{\omega}{\pi} \dashint_0^\infty 
\ln \left( \sum_{j=1}^N \mu_j+2 \sum_{j> k}^N \sqrt{\mu_j \mu_k} 
\cos (\omega' \Delta t_{jk}-\pi n_{jk}) \right) 
\frac{d\omega'}{\omega'^2-\omega^2} -i \ln B(\omega),
\end{align}
where $\Delta t_{jk}=\Delta t_j-\Delta t_k$ and $n_{jk}=n_j-n_k$.
Note that although the lower limit of integration is $\omega'=0$, 
where the geometrical optics approximation breaks down, 
we have used $F_{\rm geo} (\omega')$ over the entire range of integration. 
This is justified at leading order, as the resultant error 
is of order ${\cal O}(1/\omega)$.

Now, consider a lens signal for which the first-arrival image is not the brightest one among 
all the images (i.e., there exists $j > 1$ such that $\mu_j > \mu_1$). 
In this case, $F_{\rm geo}(\omega)$ can encircle the origin in the complex $\omega$-plane as $\omega$ increases, indicating that the phase of $F_{\rm geo}(\omega)$ can grow arbitrarily large.
On the other hand, the first term on the right-hand side of Eq.~(\ref{relation-high-f}) does not grow unboundedly 
with increasing $\omega$, due to the presence of $\omega^2$ in the denominator. 
Thus, the unbounded behavior of the left-hand side is compensated by the Blaschke product on the right-hand side.

\begin{figure}[t]
  \begin{center}
    \includegraphics[clip,width=11.0cm]{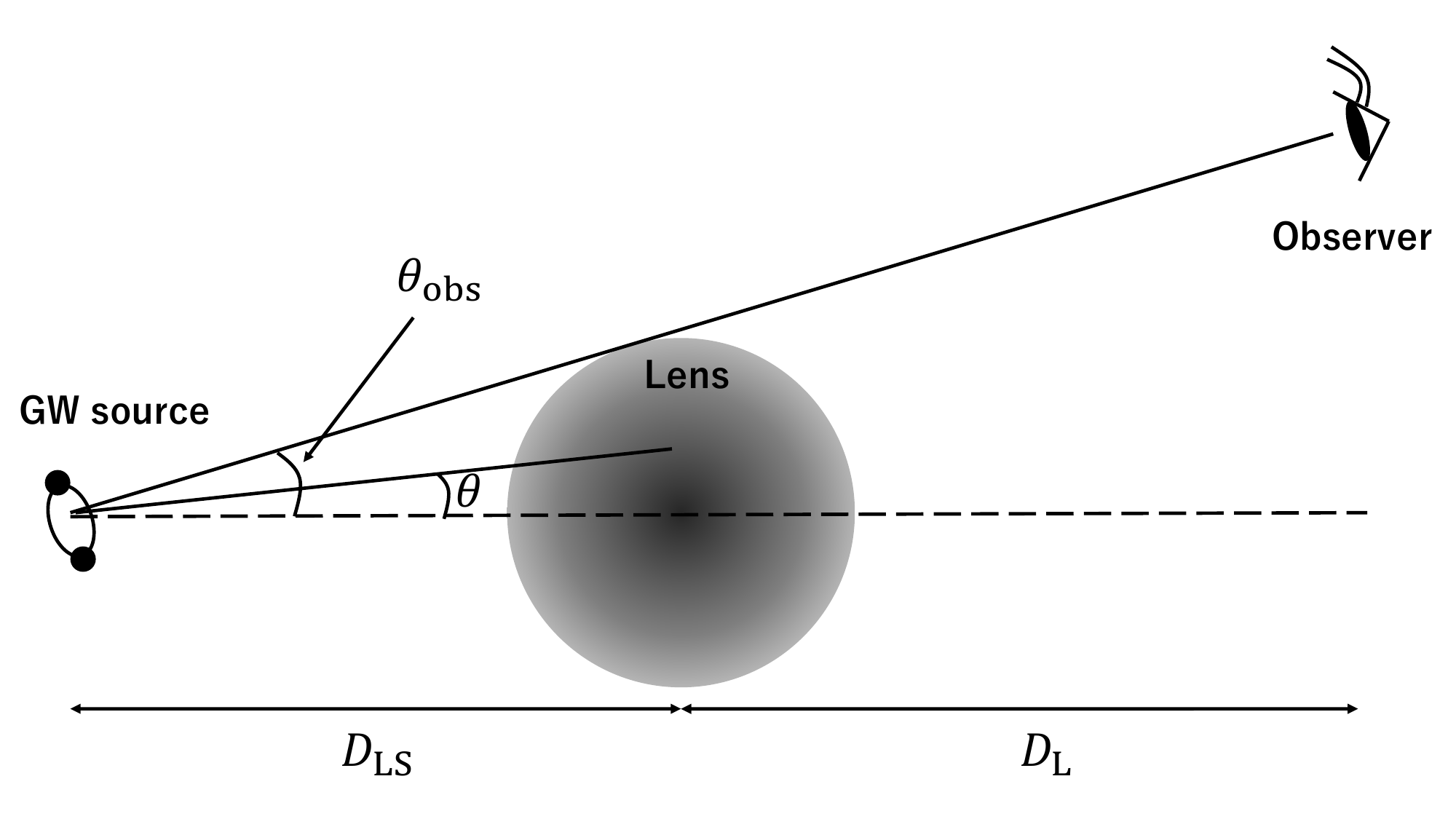}
    \caption{Schematic picture showing the configuration of the GW source, 
    lens, and the observer.}
    \label{fig:NFW-configuration}
  \end{center}
\end{figure}

To demonstrate that such behavior can occur in a realistic lens system, 
we consider the amplification factor for a lens described by the NFW profile for which the mass density profile
is given by:
\be
\rho (r)=\frac{\rho_s}{r/r_s {\left( 1+r/r_s \right)}^2},
\ee
where $\rho_s, r_s$ are constants. 
For a general lens profile, the amplification factor at the observer's position ${\bm y}$ is expressed as:
\be
F(w,{\bm y})=\frac{w}{2\pi i}\int d{\bm x}~ e^{iw t_d ({\bm x},{\bm y})} 
\ee
where $t_d ({\bm x},{\bm y})$ defined by: 
\be
t_d ({\bm x},{\bm y})= \frac{1}{2} {({\bm y}-{\bm x})}^2- \psi ({\bm x})-\phi_m ({\bm y}),
\ee
is the light travel time and $\psi ({\bm x})$ is the lens potential.
The last term $\phi_m ({\bm y})$ has been introduced so that $\min_{\bm x} t_d ({\bm x},{\bm y})=0 $.
For the NFW profile, the lens potential is given by \cite{Takahashi:dt}:
\begin{align}
\psi (x)=\left\{
\begin{array}{l}
\frac{\kappa_s}{2} \Big[ {\left( \ln \frac{x}{2} \right)}^2-{({\rm arctanh} \sqrt{1-x^2})}^2  \Big]~~~~~x\le 1  \\
\frac{\kappa_s}{2} \Big[ {\left( \ln \frac{x}{2} \right)}^2+{({\rm arctan} \sqrt{x^2-1})}^2  \Big] ~~~~~~~x >1
\end{array}
\right.
\end{align}
where $\kappa_s \equiv 16\pi G \rho_s r_s \frac{D_{\rm L} D_{\rm LS}}{D_{\rm S}}$ and:
\be
w=\frac{D_{\rm S} r_s^2}{D_{\rm L} D_{\rm LS}}\omega, ~~~x=\frac{D_{\rm LS}}{r_s} \theta,~~~y=\frac{D_{\rm LS}}{r_s} \theta_{\rm obs}
\ee
are dimensionless frequency and positions.
$D_{\rm S}$, $D_{\rm L}$, and $D_{\rm LS}$ denote the distances between the source and the observer, the lens and the observer, and the lens and the source, respectively.
The angle $\theta$ is measured from the line connecting the source and the lens, 
while $\theta_{\rm obs}$ denotes the corresponding angle at the observer 
(See Fig.~\ref{fig:NFW-configuration}). 
%\sk{Is the optical axis the line connecting the source and the lens, or the observer and the lens?}

\begin{figure}[t]
  \begin{center}
    \includegraphics[clip,width=13.0cm]{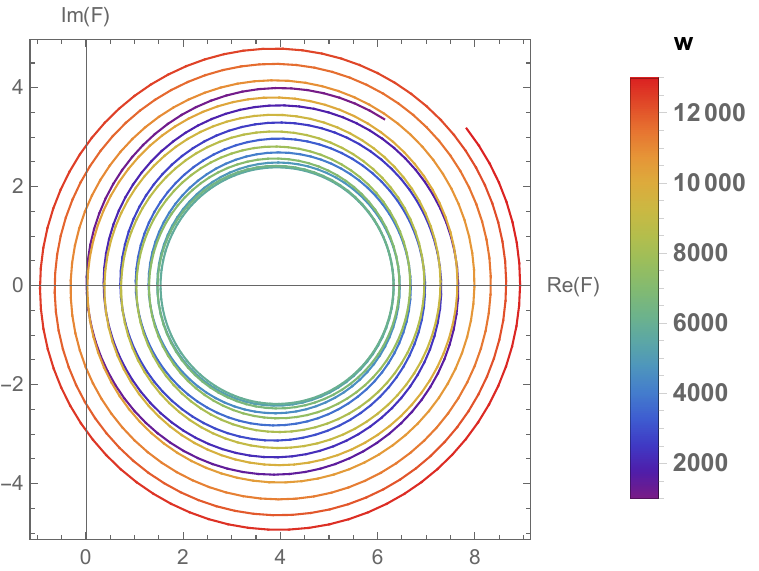}
    \caption{Trajectory of $F(\omega)$ as $\omega$ is varied for the case where the lens profile is 
    given by the NFW profile. $x/(y)$ axis is ${\rm Re} F(\omega)/({\rm Im} F(\omega))$ 
    and the range of $\omega$ in terms of $w$ is $2\times 10^3 \le w \le 1.2\times 10^4$.}
    \label{fig:NFW-amplification-factor}
  \end{center}
\end{figure}

Fig.~\ref{fig:NFW-amplification-factor} shows the trajectory of $F(\omega)$, 
computed in the geometrical optics approximation, 
in the complex plane as $w$ is varied over the range $10^3 \le w \le 1.2 \times 10^4$. 
The parameters are fixed at $\kappa_s = 1$ and $y = 0.025$, for which three images are formed 
with magnifications $(\mu_1, \mu_2, \mu_3) = (15.40, 17.02, 3.02)$. 
As $\omega$ increases, $F(\omega)$ moves in the counter-clockwise direction and passes through 
regions with negative ${\rm Re}F(\omega)$, 
providing clear evidence that the phase of $F(\omega)$ grows without bound.
The unlimited growth of the phase for the NFW profile is also seen in \cite{Takahashi:dt}.

The above consideration shows that the phase of $F(\omega)$ cannot be uniquely determined
from the magnitude of $F(\omega)$ in general due to the possible existence of the contribution
from the Blaschke product:
\be
\label{KK:phase-from-magnitude}
\theta (\omega) =-\frac{2\omega}{\pi} \dashint_0^\infty
\frac{\ln |F(\omega')|}{\omega'^2-\omega^2}d\omega'-i \sum_n \ln \left( \frac{\omega-\mu_n}{\mu_n^*-\omega} \right).
\ee
This means that measurement of the magnitude of $F(\omega)$ alone does not fix the phase of $F(\omega)$
and additional assumption has to be made to completely determine the phase.
On the other hand, by substituting $\omega_1=0, ~\omega_2=\omega$ to 
Eq.~(\ref{magnitude-from-phase}) and using the fact $F(0)=1$,
the magnitude of $F(\omega)$ is uniquely determined from the phase of $F(\omega)$ as: 
\be
\label{KK:magnitude-from-phase}
\ln |F(\omega)|=\frac{2}{\pi} \dashint_0^\infty \left(
\frac{\omega'}{\omega'^2-\omega^2}-\frac{1}{\omega'} \right) \theta (\omega')
d\omega'.
\ee
%\sk{Is the crux of this section that given the phase of F, the amplitude is uniquely determined. Conversely, given the amplitude of F, the phase is not uniquely determined because of the Blaschke product. Is this correct? I also don't fully understand why $\omega_1 = 0, \omega_2 = \omega$. Is this a requirement of the KK relation? }
%\ts{Yes, that is correct. As for the choice $\omega_1=0$, 
%because Eq.~(\ref{magnitude-from-phase}) contains $|F(\omega)|$ at two different frequencies, 
%we need to fix one of them to translate into a relation which gives $|F|$ at a specific frequency
%$\omega$. The choice $\omega_1=0$ works well for this purpose as $F(0)=1$ for any %lens model.
%In principle, we can choose a different value for $\omega_1$ if $F(\omega_1)$ is known a priori by some reason.}

\subsection{Numerical demonstration of the KK relation}
Although it has been established that the KK relations (\ref{KK:phase-from-magnitude}) and 
(\ref{KK:magnitude-from-phase}) hold true for amplification factor of any lens system,
it may be instructive to numerically confirm the validity of the relations for some simple lens system.
In this subsection, we consider a point-mass lens as an example and numerically compare the
left-hand side and the right-hand side of the relation (\ref{KK:phase-from-magnitude}).

There exits an analytic form of the amplification factor for the point-mass lens 
and it is given by: \cite{schneider2012gravitational}
\be
\label{F:PML}
F(\omega)=\exp \bigg[ \frac{\pi w}{4}+\frac{iw}{2} \left( \ln \left( \frac{w}{2}\right)-2\tau_{\rm min} \right)\bigg] \Gamma \left( 1-\frac{iw}{2}\right) {}_1F_1 \left( \frac{iw}{2},1;\frac{iwy^2}{2} \right),
\ee
where $w\equiv 4GM\omega$, $M$ is the lens mass, $y$ is the impact parameter normalized 
by the Einstein radius, and $\tau_{\rm min}$ is defined by:
\be
\tau_{\rm min}(y)=\frac{2}{{(y+\sqrt{y^2+4})}^2}
-\ln \left( \frac{y+\sqrt{y^2+4}}{2} \right).
\ee

\begin{figure}[t]
  \begin{center}
    \includegraphics[clip,width=10.0cm]{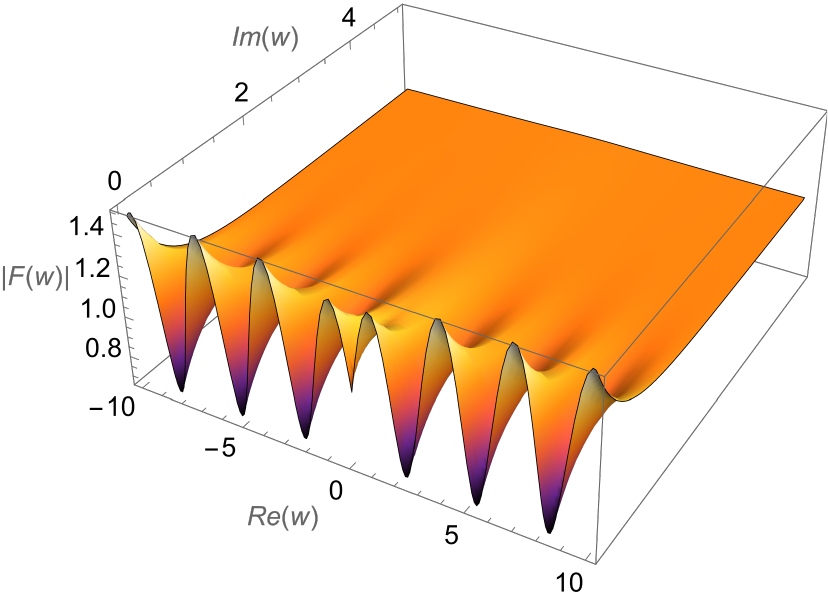}
    \caption{Plot of $|F(\omega)|$ in the upper-half complex $\omega$-plane for the point-mass lens with $y=1$. }
    \label{fig:point-mass-abs-F}
  \end{center}
\end{figure}

Fig.~\ref{fig:point-mass-abs-F} shows $|F(\omega)|$ in the upper-half complex $\omega$-plane for the point-mass lens with $y=1$.
We find that there are no points in the upper-half complex plane where $F(\omega)$ vanishes.
Thus, Blaschke product is absent (i.e. $B=1$) in this case 
and $\ln F(\omega)$ is analytic in the upper-half complex plane.
This is also consistent with the fact that in geometrical optics two images appear for the point-mass lens 
and the first-arrival image has the largest magnification.

\begin{figure}
\begin{center}
\begin{tabular}{cc}
\begin{minipage}[t]{0.5\hsize}
\includegraphics[width=80mm]{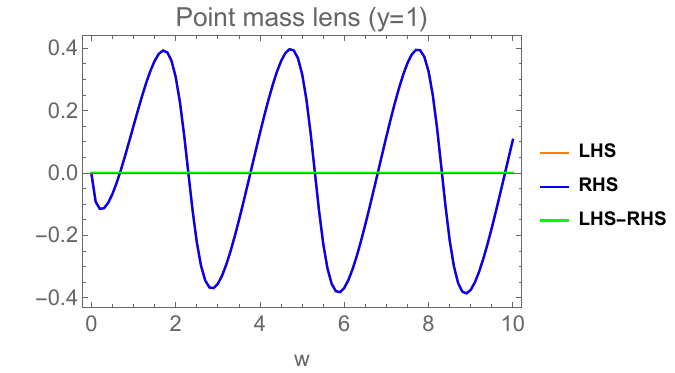}
\end{minipage}
\begin{minipage}[t]{0.5\hsize}
\includegraphics[width=63mm]{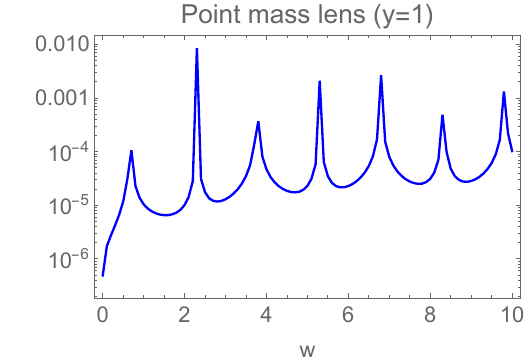}
\end{minipage}
\end{tabular}
\caption{The left panel shows a plot of the left-hand side and the right-hand side of the KK relation (\ref{KK:phase-from-magnitude}) as a function of $w=4 GM\omega$ ($y$ is fixed to $y=1$). 
The right panel shows the relative error between the two sides. }
\label{fig:PML-comparison-yfixed}
\end{center}
\end{figure}

\begin{figure}
\begin{center}
\begin{tabular}{cc}
\begin{minipage}[t]{0.5\hsize}
\includegraphics[width=79mm]{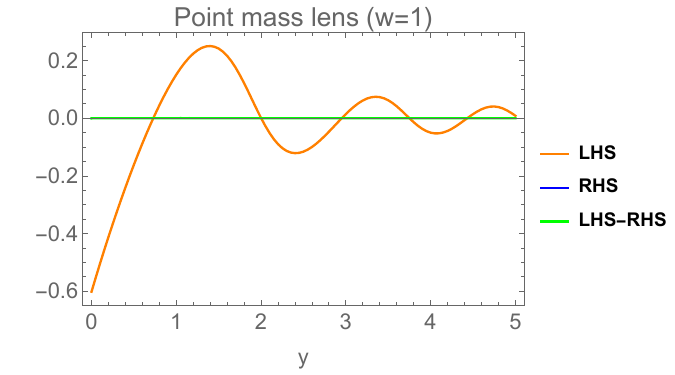}
\end{minipage}
\begin{minipage}[t]{0.5\hsize}
\includegraphics[width=63mm]{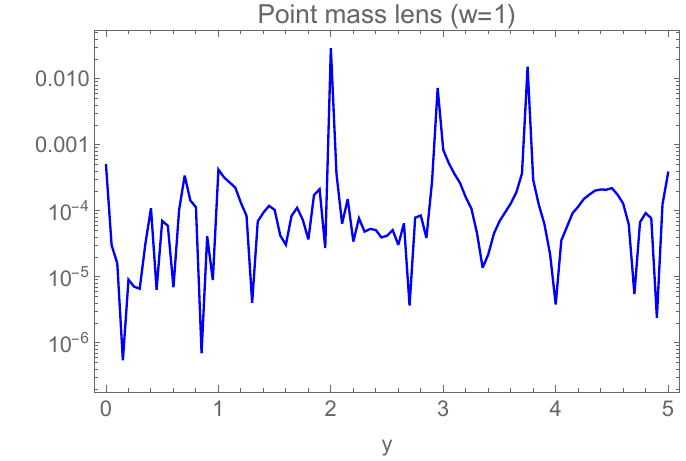}
\end{minipage}
\end{tabular}
\caption{The left panel shows a plot of the left-hand side and the right-hand side of the KK relation (\ref{KK:phase-from-magnitude}) as a function of $y$ ($w$ is fixed to $w=1$). 
The right panel shows the relative error between the two sides. }
\label{fig:PML-comparison-wfixed}
\end{center}
\end{figure}

The left panel of Fig.~\ref{fig:PML-comparison-yfixed} compares the left-hand and right-hand sides 
of the KK relation (\ref{KK:phase-from-magnitude}) for a point-mass lens with $B=1$, 
plotted as functions of $w$ at fixed $y=1$. 
The two curves visually overlap almost perfectly. 
The right panel shows the relative error between the two sides. 
While the error is nonzero for all $w$, it arises from a numerical artifact: 
the upper limit of the integral on the right-hand side of (\ref{KK:phase-from-magnitude}) is 
truncated at finite frequency. 
The relative error increases near frequencies where the phase vanishes, 
which explains the oscillatory pattern.
Fig.~\ref{fig:PML-comparison-wfixed} presents a similar comparison, 
this time as a function of $y$ with $w$ fixed at $w=1$. 
The qualitative behavior is consistent with that in Fig.~\ref{fig:PML-comparison-yfixed}. 
Overall, this analysis offers numerical validation of the KK relation in the point-mass lens system.
%\sk{From a data analysis perspective, an inspection by eye should be complemented by an actual calculation of the relative error (abs(difference)/true value).}

\section{Universal features at low-frequency regime}
The KK relations (\ref{KK:phase-from-magnitude}) and (\ref{KK:magnitude-from-phase})
for the amplification factor relate one of its magnitude and phase to the other.
Because they contain integration over frequency, 
knowledge of the magnitude/(phase) over all the frequency range is necessary
to determine the phase/(magnitude) at a specific frequency.
However, as we will demonstrate below, in the low-frequency regime,
leading order behavior of $|F(\omega)|-1$ completely determines
the leading order term of the phase $\theta (\omega)$.

To this end, let us suppose that $|F(\omega)|$ for sufficiently small $\omega$
is given by:
\be
\label{low-f:magnitude}
|F(\omega )|=1+A \omega^\alpha +\cdots,
\ee
where $A$, which depends on the lens profile, is constant independent of $\omega$
and $\alpha$, which also depends on the lens profile, is another constant in the range $0<\alpha \le 1$.
It is known that $\alpha=1$ for point-mass lens, $\alpha=\frac{1}{2}$ for
singular isothermal shere (SIS) profile,
and $\alpha=\frac{k}{2}$ for generalized SIS profile for which $\rho (r) \propto r^{-k-1}$ \cite{Choi:2021bkx, Tambalo:2022plm}.
Terms higher order in $\omega$ are denoted by $\cdots$ and they are ignored in our computations.

Let us evaluate the phase in the low-frequency regime by substituting the 
expansion (\ref{low-f:magnitude}) into Eq.~(\ref{KK:phase-from-magnitude}).
We first focus on the first term on the right-hand side of Eq.~(\ref{KK:phase-from-magnitude}).
For convenience, we rewrite it as:
\be
\label{low-f:first-term}
-\frac{2\omega}{\pi} \dashint_0^\infty \frac{\ln |F(\omega')|}{\omega'^2-\omega^2}d\omega'
=
-\frac{2\omega}{\pi} \int_0^\infty \frac{\ln |F(\omega')|-\ln |F(\omega)|}{\omega'^2-\omega^2}d\omega'.
\ee
This identity holds due to the following identity:
\be
\dashint_0^\infty \frac{1}{\omega'^2-\omega^2}d\omega'
=0.
\ee
In the new expression, evaluating the Cauchy principal value is not necessary
because the numerator $\ln |F(\omega')|-\ln |F(\omega)|$ vanishes at $\omega'=\omega$.

To compute the integral, we consider the case $\alpha <1$.
The case with $\alpha=1$ will be treated separately.
From the structure of the integrand, it is evident that the contribution from the region 
$\omega' \gg \omega$ is convergent. 
Consequently, this region contributes at order ${\cal O}(\omega)$ to Eq.~(\ref{low-f:first-term}).
On the other hand, contribution coming from the region $\omega'={\cal O}(\omega)$ can be evaluated as:
\be
\label{low-f:integral-alpha}
-\frac{2\omega}{\pi} \dashint_0^\infty \frac{\ln |F(\omega')|}{\omega'^2-\omega^2}d\omega' =-\frac{2\omega A}{\pi} \int_0^\infty \frac{\omega'^\alpha-\omega^\alpha}{\omega'^2-\omega^2}d\omega' =-A \tan \left( \frac{\pi \alpha}{2} \right) \omega^\alpha.
\ee
Here we have changed the integration variable from $\omega'$ to $s$ 
by $\omega'=\omega s$ and used the formula $\int_0^\infty \frac{s^\alpha-1}{s^2-1}ds
=\frac{\pi}{2}\tan \left( \frac{\pi \alpha}{2}\right)$.  
Since $\alpha <1$, this gives the leading order contribution to Eq.~(\ref{low-f:first-term}) \footnote{
In Eq.~(\ref{low-f:integral-alpha}), the upper limit of the integral has been pushed to $\infty$ where
the low-frequency expansion (\ref{low-f:magnitude}) breaks down.
However, the error caused by this procedure is ${\cal O}(\omega)$ and does not affect our conclusion
that Eq.~(\ref{low-f:integral-alpha}) is the leading order term.
}.
If $\alpha=1$, Eq.~(\ref{low-f:first-term}) diverges and the use of the low-frequency expansion 
(\ref{low-f:magnitude}) up to $\omega' \to \infty$ is not allowed to derive the leading order term.
Thus, the upper limit of integral $\Lambda$ should be in the low-frequency regime and
the integral can be evaluated as:
\be
-\frac{2\omega A}{\pi} \int_0^\Lambda \frac{\omega'-\omega}{\omega'^2-\omega^2}d\omega'
=-\frac{2\omega A}{\pi}\ln \left( \frac{\Lambda+\omega}{\omega} \right) =\frac{2A}{\pi}\omega \ln \omega +\cdots
\ee
where $\omega \ll \Lambda$ is used in the last step.

As for the second term on the right-hand side of Eq.~(\ref{KK:phase-from-magnitude}),
noting that it vanishes in the limit $\omega \to 0$, the leading order contribution is given by:
\be
-i \sum_n \ln \left( \frac{\omega-\mu_n}{\mu_n^*-\omega} \right)
=2 \left( \sum_n \frac{{\rm Im}~ \mu_n}{{|\mu_n|}^2} \right) \omega +\cdots.
\ee
Thus, the Blaschke product gives contribution only at the order ${\cal O}(\omega)$ and 
can be ignored as long as the leading order contribution is concerned.

To summarize, at the leading order, the phase of $F(\omega)$ is written in terms of the
magnitude of $F(\omega)$ as: 
\begin{align}
\label{low-f-alpha:phase-from-mag}
\theta (\omega)=\left\{
\begin{array}{l}
-\tan \left(  \frac{\pi \alpha}{2} \right) \ln |F(\omega)|+\cdots~~~~~0\le \alpha \le 1  \\
\ln |F(\omega)| \ln \omega +\cdots \hspace{22mm} \alpha =1
\end{array}
\right.
\end{align}
In the literature \cite{Choi:2021bkx, Tambalo:2022plm}, low-frequency behaviors of $F(\omega)$ for 
some representative lens profiles are obtained as:
\begin{align}
F(\omega) =\left\{
\begin{array}{l}
1+2^{-\frac{k}{2}} e^{-i\frac{k\pi}{4}} \Gamma \left( 1-\frac{k}{2} \right) w^\frac{k}{2}+\cdots 
~~~~~{(\rm Generalized~SIS)} \\
1+\frac{w}{2} \left( \pi+2i \ln w \right)+\cdots \hspace{18mm} {(\rm Point-mass ~lens)}
\end{array}
\right.
\end{align}
Using the correspondence $k=2\alpha$, it can be verified that the leading order behavior
of these amplification factors satisfies the relation (\ref{low-f-alpha:phase-from-mag}).
This demonstrates that the low-frequency behavior of $F(\omega)$ is strongly restricted by the
KK relation which originates from the causal nature of gravitational lensing.

\section{Construction of phase from magnitude}
As we have shown in Sec.~\ref{Sec:Derivation-KK}, the phase of $F(\omega)$ cannot be uniquely
determined from the magnitude $|F(\omega)|$ due to potential contribution from the Blaschke product.
However, for the case of weak lensing for which $F(\omega)$ deviates only slightly from unity or some
lens systems such as the point-mass lens for which ${\rm Re}F(\omega)$ never crosses negative region,
the Blaschke product is $B=1$ and the phase is uniquely fixed from the magnitude.

In \cite{Chakraborty:2024mbr}, inspired by the shape of $F(\omega)$ for some simple lens models,
a following phenomenological fitting form of $|F(\omega)|$ has been proposed:
\begin{equation}
\label{pheno:Chakraborty}
    |F(\omega)| = a + b e^{-k\omega/\omega_0}\cos (\omega /\omega_0 +\varphi),
\end{equation}
where $(a, b, k,\omega_0, \varphi )$ are fitting parameters.
Since the physical condition $F(0)=1$ fixes one of the parameters, independent parameters are four.
In this section, as an application of the KK relation derived in this paper,
we will construct the phase of $F(\omega)$ whose magnitude is given by Eq.~(\ref{pheno:Chakraborty})
under the assumption that the Blaschke product is absent. 

Using the equivalent expression (\ref{low-f:first-term}), the phase can be written as:
\be
\label{phase:Chakraborty}
\theta (\omega)=-\frac{2\omega}{\pi \omega_0} \int_0^\infty 
\frac{\ln |F(\omega_0s)|-\ln |F(\omega)|}{s^2-\frac{\omega^2}{\omega_0^2}} ds,
\ee
where the integration variable has been changed to $\omega' \to s=\omega'/\omega_0$.
For the form (\ref{pheno:Chakraborty}), it is evident that the above phase only depends on the 
combination $\omega /\omega_0$.

\begin{figure}
\begin{center}
\begin{tabular}{cc}
\begin{minipage}[t]{0.5\hsize}
\includegraphics[width=7cm]{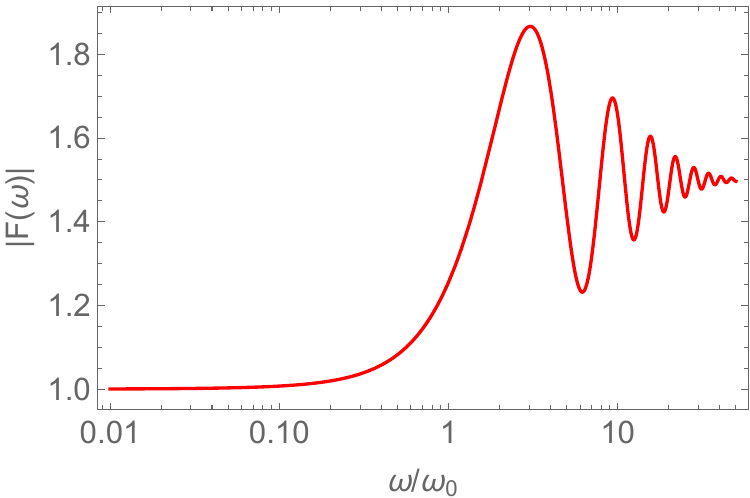}
\end{minipage}
\begin{minipage}[t]{0.5\hsize}
\includegraphics[width=7cm]{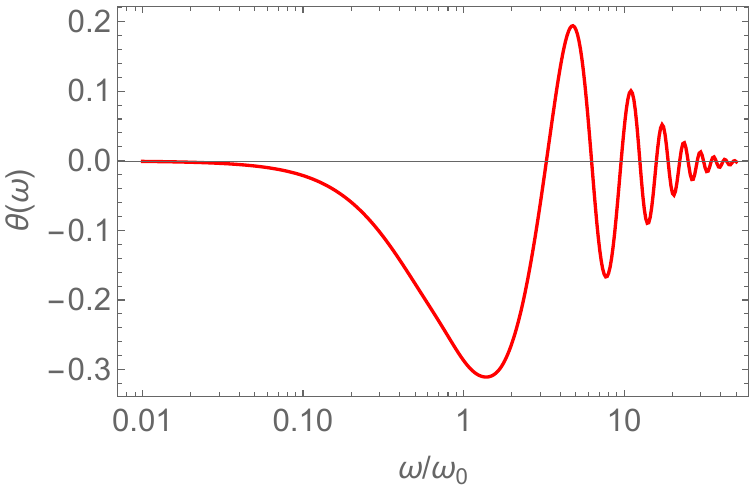}
\end{minipage}
\end{tabular}
\caption{Left panel: plot of $|F(\omega)|$ given by Eq.~(\ref{pheno:Chakraborty}) as a function of 
$\omega /\omega_0$ for a parameter choice
$(b,k,\varphi)=(0.5,0.1,\pi)$. 
Right panel: plot of the phase $\theta (\omega)$ computed by the KK relation (\ref{phase:Chakraborty}) where the magnitude $|F(\omega)|$ is given by the left panel.}
\label{fig:Pheno-form}
\end{center}
\end{figure}

Fig.~\ref{fig:Pheno-form} shows $|F(\omega)|$ given by Eq.~(\ref{pheno:Chakraborty}) as a function of 
$\omega /\omega_0$ for a parameter choice
$(b,k,\varphi)=(0.5,0.1,\pi)$ (left panel).
The right panel is a plot of $\theta (\omega)$ of $F(\omega)$ whose magnitude is given by the left panel,
which is obtained by computing Eq.~(\ref{phase:Chakraborty}).
On low-frequency side, it is seen that $\theta (\omega)$ decreases as $\omega$ is increased.
This is consistent with the low-frequency behavior given by Eq.~(\ref{low-f-alpha:phase-from-mag})
(in the present case, $\alpha=1$).
In this way, if a phenomenological form for either $|F(\omega)|$ or $\theta (\omega)$ is proposed,
the other cannot be independent but is fixed by the KK relation.

It may be useful to have a phenomenological form for $|F(\omega)|$ or $\theta (\omega)$
such that the other can be computed analytically using the KK relation.
To this end, we provide a simple expression that satisfies this requirement.
It is given by:
\be
\label{analytic:lnF}
\ln |F(\omega)|=\frac{\alpha \omega}{\omega+\beta} \cos (\gamma \omega +\phi),
\ee
where $\alpha (>0), \beta (>0), \gamma (>0)$, and $\phi$ are fitting parameters. 
This form is designed so that, at large frequency, it asymptotes to the oscillating function characteristic of geometrical optics, 
and at small frequencies, it scales as $\propto \omega$.
In fact, the asymptotic behaviors of $\ln F$ can be understood 
immediately:
\begin{align}
\ln |F(\omega)| =\left\{
\begin{array}{l}
\frac{\alpha}{\beta}\omega \cos \phi +{\cal O}(\omega^2)
\hspace{22mm}  {(\rm for~small~\omega)} \\
\alpha \cos (\gamma \omega +\phi) +\cdots \hspace{18mm} {(\rm for~large ~\omega)}
\end{array}
\right.
\end{align}

\begin{figure}
\begin{center}
\begin{tabular}{cc}
\begin{minipage}[t]{0.5\hsize}
\includegraphics[width=79mm]{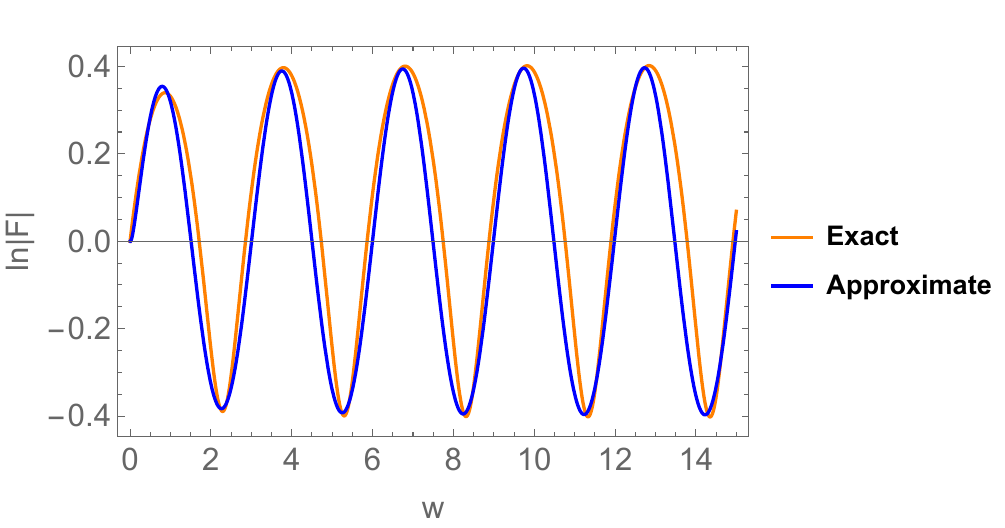}
\end{minipage}
\begin{minipage}[t]{0.5\hsize}
\includegraphics[width=80mm]{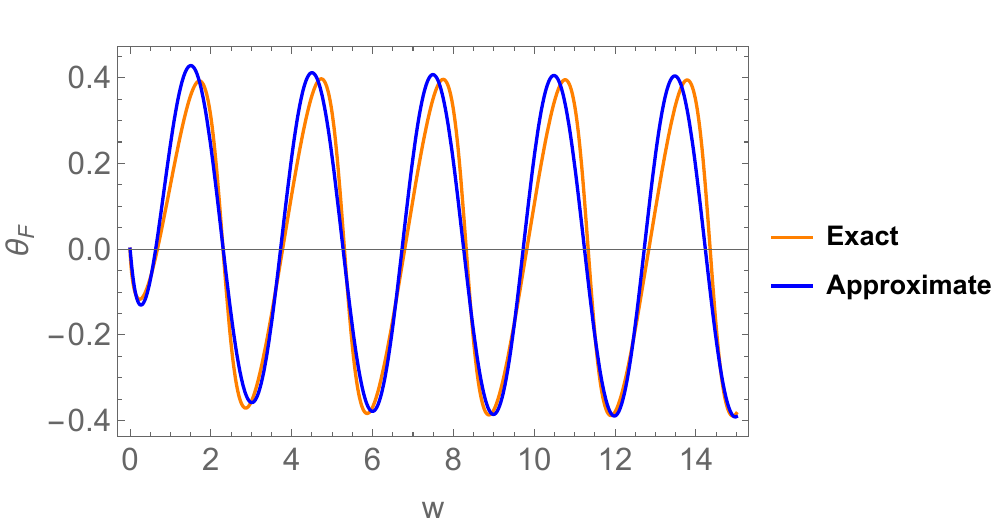}
\end{minipage}
\end{tabular}
\caption{The left panel shows a comparison between $\ln |F(w)|$ for the point-mass lens with $y=1$ (Exact) and Eq.~(\ref{analytic:lnF}) 
with $(\alpha, \beta, \gamma, \phi)=
(0.4,0.1,2.1,-1.6)$ (Approximate). 
The right panel shows a comparison between $\arg F (\omega)$ 
for the point mass and $\theta (\omega)$ given by Eq.~(\ref{analytic:phase}) 
for the same values of the parameters as in the left panel.}
\label{fig:analytic}
\end{center}
\end{figure}

The left panel of Fig.~\ref{fig:analytic} shows a comparison between $\ln |F(w)|$ for the point-mass lens
(see Eq.~(\ref{F:PML})) with $y=1$ and Eq.~(\ref{analytic:lnF}) with $(\alpha, \beta, \gamma, \phi)=
(0.4,0.1,2.1,-1.6)$.
The values of the parameters $(\alpha, \beta, \gamma, \phi)$ have been chosen to provide a good fit to the point-mass lens. 
It is evident that the phenomenological form (\ref{analytic:lnF}) captures the essential behaviors of the
physical amplification factor, at least in the case of the
point-mass lens.
%present case \sk{What is meant by present case?}.

For $\ln |F(\omega)|$ given by Eq.~(\ref{analytic:lnF}),
the integral appearing on the right-hand side of Eq.~(\ref{KK:phase-from-magnitude})
can be performed analytically and the result is given by:
\begin{align}
\label{analytic:phase}
\theta (\omega)=& -\frac{\alpha \, \omega}{\pi (\beta - \omega)(\beta + \omega)} \Big[ 
-2\beta \cos(\beta \gamma - \phi) \, \text{Ci}(\beta \gamma) \nonumber \\
&+ 2 \, \text{Ci}(\gamma \omega) \left( \beta \cos\phi \cos(\gamma \omega) 
+ \omega \sin\phi \sin(\gamma \omega) \right) + \beta \sin(\beta \gamma - \phi) \left( \pi - 2 \, \text{Si}(\beta \gamma) \right) \nonumber \\ 
&+ \cos\phi \sin(\gamma \omega) \left( -\pi \omega + 2\beta \, \text{Si}(\gamma \omega) \right)  + \cos(\gamma \omega) \sin\phi \left( \pi \beta - 2\omega \, \text{Si}(\gamma \omega) \right)
\Big],
\end{align}
where: 
\be
\mathrm{Ci}(z) = - \int_z^{\infty} \frac{\cos t}{t} \, dt,~~~~~
\mathrm{Si}(z) = \int_0^{z} \frac{\sin t}{t} \, dt
\ee
are Cosine and Sine Integral, respectively.
The right panel of Fig.~\ref{fig:analytic} shows a comparison between
$\arg F(\omega)$ for the point mass and $\theta (\omega)$ given
by Eq.~(\ref{analytic:phase}) for the same values of the parameters as
in the left panel.
Overall, the two curves mostly overlap, explicitly demonstrating that 
once a phenomenological and
physically plausible form of $\ln |F(\omega)|$ is provided as an approximation to a certain physical amplification factor,
the phase computed using the KK relation remains close to the phase
of the physical amplification factor.

\section{Conclusions}
In this work, we have reformulated the Kramers-Kronig (KK) relation 
for the amplification factor into a relation between the magnitude $|F(\omega)|$ and the phase $\theta(\omega)=\arg F(\omega)$. 
While the original KK relation connects the real and imaginary parts of $F(\omega)$, our reformulation offers a more direct connection to quantities which
have clear physical meaning.

A key insight from our analysis is that the phase $\theta(\omega)$ is not 
always uniquely determined by the magnitude $|F(\omega)|$, 
due to the possible presence of a Blaschke product. 
This ambiguity, absent in simple cases like the point-mass lens, 
may appear in more complex lens models such as NFW density profile.
Our reformulation thus clarifies the conditions under which such degeneracies 
may arise and suggests caution when interpreting phenomenological models based solely on magnitude.

We have shown that, in the low-frequency regime, the leading-order behavior 
of the phase is completely fixed by the leading-order behavior of the magnitude. This reproduces known results for specific lens models and, 
more importantly, establishes a model-independent constraint derived 
solely from the KK relation. 
This universal constraint will be of practical value in testing the consistency 
of observed lensing features in GW signals.

As further applications, we studied two examples where the phase is 
derived from phenomenological and analytic forms of the magnitude 
using the KK relation. 
Interestingly, in the second case, the KK integral can be evaluated analytically, 
yielding a closed-form expression for the phase. 
These results provide a useful way to 
incorporate phase information into phenomenological modeling of lensing signals.

Our results show that the magnitude and phase of the amplification factor are not independent quantities, but are tightly connected by the KK relation.
These should be taken into account in detecting and characterizing 
lensing signals in GW data, 
especially in agnostic or phenomenological approaches.

\section*{Acknowledgments}
This work was supported by JSPS KAKENHI Grant Number JP23K03411 (TS). S.J.K. gratefully acknowledges support from SERB grants SRG/2023/000419 and MTR/2023/000086.

\bibliography{draft}

\end{document}